# Numerical and experimental analyses of resin infusion manufacturing processes of composite materials

P.WANG[1], S.DRAPIER[1], J.MOLIMARD[1], A.VAUTRIN[1], J.C. Minni[2]

[1] Mechanics and Materials Processing Dep., Structures and Materials Science Division

and Laboratory for Tribology and Systems Dynamics, UMR CNRS 5513

École des Mines de Saint Etienne 42023 Saint-Étienne Cedex 02, France

[2] Hexcel Corporation SAS, 38630 Les Avenières, France

**Abstract:** Liquid Resin Infusion (LRI) processes are promising manufacturing routes to produce large, thick or complex structural parts. They are based on the resin flow induced, across its thickness, by a pressure applied onto a preform / resin stacking. However, both thickness and fiber volume fraction of the final piece are not well controlled since they result from complex mechanisms which drive the transient mechanical equilibrium leading to the final geometrical configuration. In order to optimize both design and manufacturing parameters, but also to monitor the LRI process, an isothermal numerical model has been developed which describes the mechanical interaction between the deformations of the porous medium and the resin flow during infusion [1,2]. With this numerical model, it is possible to investigate the LRI process of classical industrial part shapes. To validate the numerical model, first in 2D, and to improve the knowledge of the LRI process, the present study details a comparison between numerical simulations and an experimental study of a plate infusion test carried out by LRI process under industrial conditions. From the numerical prediction, the filling time, the resin mass and the thickness of the preform can be determined. On another hand, the resin flow and the preform response can be monitored by experimental methods during the filling stage. One key issue of this research work is to highlight the changes in major process parameters during the resin infusion stage, such as the temperature of the preform and resin, and the variations of both thickness and fiber volume fraction of the preform. Moreover, this numerical / experimental approach is the best way to improve our knowledge on the resin infusion processes, and finally, to develop simulation tools for the design of advanced composite parts.

**Key words:** Liquid Resin Infusion, numerical simulation, comparison, industrial condition, resin flow, filling time, thickness, fiber volume fraction.

## 1. Introduction

During the last decade, the Resin Infusion Processes (RIP) have become popular for manufacturing structural polymer-based composites. RIP have been indentified as cost-effective alternative to conventional autoclave manufacturing technique. For example, with RIP it is possible to produce complex and thick parts with very good mechanical properties and with less waste than traditional methods [3, 4]. However, the process is rather difficult to control, first because the mechanisms driving the infusion stage are quite complex, and second with the existing industrial technology physical parameters such as thickness or resin front on small dimensions are not accessible. Since industrially the thickness must be controlled precisely, understanding in details the filling stage of infusion is of prime importance.

As one type of RIP, Liquid Resin Infusion (LRI) process seems quite promising. In this process (Fig.1), resin is distributed through a highly permeable flow enhancement fabric placed on top of the fibres perform stacking. Due to a pressure differential created by a vacuum at the vent of the system, resin impregnates across the compressible preforms, *i.e.* in the direction transverse to the preform 'plane'. The LRI process leads to final part quality improvement since the resin filling and curing stages are distinct. On the contrary, the thickness and fiber volume fraction of the final piece are not well controlled during the process because, first, of the use of a vacuum bag instead of a rigid mould and second, due to the large preform deformation when vacuum and pressure are applied. Therefore, the final properties of the composite parts strongly depend on the process parameters. In order to optimize both design and manufacturing parameters, a numerical model has been developed which describes the mechanical interaction between the deformation of the preform and the resin flow during infusion stage [1]. To validate the numerical model and to improve the knowledge of the resin infusion process, this research work will deal with the numerical simulations and experimental studies of the major process parameters of a plate infusion test carried out by LRI process under industrial conditions.

Figure 1 about here

## 2. Resin infusion modelling

Early numerical model of resin infusion processes developed can be found for example in the work of *Loos and MacRae* [5]. Authors developed a two-dimensional analytical model for Resin Film Infusion (RFI) process, which takes into account the porosity and compaction of the vacuum bag, but they did not study the resin-preforms interaction during the deformation of the preform. Then, *Ambrosi and Preziosi* [6] proposed an approach to deal with the injection processes in elastic porous preform for one-dimensional problems by using a modified momentum balance equation of the fluid and solid phases.

Recently, several models have been developed for the resin flow and the response of the preform in Liquid Composite Moulding (LCM) processes [7-13, 26-28]. Generally, the models mentioned provide some partial information, but they are not suitable for integration into solvers under our industrial conditions, based on the finite element method.

More recently, an exhaustive model has been developed by *Celle et al.* [1, 2]. It was established and implemented in an industrial environment by coupling general 3D formulations of solid, fluid, and porous mechanics to represent a transient resin flow in an isothermal compressible porous medium. It is based on the resin flow induced across its thickness by pressure applied onto a preform / resin stacking. A strong coupling between resin flow and response of the preform was proposed in this model. The implementation of this model was realized by using Pro-Flot libraries and the filling algorithm of the PAM-RTM$^{TM}$ software.

## 2.1 Model geometry

In the macroscopic model from *Celle et al.* [1, 2] the two components (resin and preforms) are represented in 3 different areas separated by moving boundaries (see Fig.2). This model includes proper boundary conditions and continuity conditions at moving interfaces. The macroscopic modelling achieves a direct numerical coupling of the fluid and the solid parts while offering reasonable computation costs.

Figure 2 about here

## 2.2 Modelling the fluid part

Resin infusion processes are characterized by a very low infusion velocity. The Reynolds number measured in these processes indicates that the resin flow must be of laminar type. Classically, the resin can be considered as a Newtonian incompressible fluid [14]. Then, the constitutive law associated with this fluid can be described under the current material configuration $\bar{x}$ and at time $t$ as the following equation:

$$\bar{\bar{\sigma}}(x,t) = 2\eta \bar{\bar{D}}(x,t) - p(x,t)\bar{\bar{I}} \tag{1}$$

with $\bar{\bar{\sigma}}(x,t)$ the Cauchy stress tensor, $\bar{\bar{D}}(x,t)$ the strain rate tensor, $\eta$ the fluid dynamic viscosity, $p(x,t)$ the hydrostatic pressure in the porous medium and $\bar{\bar{I}}$ the second-order identity tensor.

### *In the purely fluid region*

A pure fluid resin area is present in the RFI process (see Fig. 2a). The resin flow is modelled in this zone by using the mass and momentum balance equations. Finally, the Stokes's flow (eq.2) is described by:

$$\begin{aligned} \eta \bar{\bar{\Delta}} \bar{v} - \bar{\nabla} p &= 0 \\ div(\bar{v}) &= 0 \end{aligned} \tag{2}$$

with $\bar{v}$ the resin velocity.

### *Resin flow within the preform*

The resin flow through the preforms consists in analysing the problem of a viscous fluid flowing in a compressible porous medium. Under a macroscopic approach, the Darcy's law (eq.3) or Brinkman's equation can describe this resin flow. We are more interested in Darcy's law because of the low permeability of our preforms (typically $10^{-11}$-$10^{-13}$ m$^2$).

$$\bar{v} = -\frac{\bar{\bar{K}}}{\eta} \cdot (\bar{\nabla} p - \rho \bar{g}) \qquad (3)$$

where $\bar{v}$ describes the Darcy's velocity, $\bar{\bar{K}}$ the permeability tensor, $p$ the resin pressure, $\rho$ the resin density and $\bar{g}$ the acceleration vector due to gravity. Moreover the local resin velocity $\bar{v}_r$ can be deduced from the Darcy's velocity $\bar{v}$ and the porosity of the preform $\phi$ ($\bar{v}_r = \bar{v}/\phi$). It is must be pointed out that permeability of the preforms is one of the main factor controlling the resin flow within the preforms. As such, it is a key parameter to modeling resin infusion [15, 16].

*Resin flow within the distribution medium*

To study the resin flow within the distribution medium (draining fabric in our cases) during LRI processes (fig.2-b), different ways are possible:

1. Model as a pure resin region,

2. Use the Brinkman's equation due to a high permeability of draining fabric,

3. Consider the approach proposed by Ngo and Tamma [17], which describes a combination of the Stokes's and the Brinkman's flow by a computational parameter $\alpha$, which equates 1 in the intra-tow region and 0 in the inter-tow or open region.

In order to simplify the numerical model, the draining fabric is represented as purely fluid region and the flow can be modelled through a Stokes approach (eq.1).

**2.3 Modelling the solid part**

Modelling the solid part focuses on the behaviour of dry and wet preforms, which can be regarded as a same solid medium. An updated Lagrangian formulation is adopted to describe this porous medium deformation. During the infusion stage, the resin hydrostatic pressure influences the response of the preform. In order to account for the resin - preform interaction, the Terzaghi's model is adopted (eq.5) [18], which takes into account directly the presence of the resin in the deforming preform through its hydrostatic pressure:

$$\overline{\overline{\sigma}} = \overline{\overline{\sigma_{ef}}} - s p_r \overline{\overline{I}} \tag{5}$$

This model postulates that the total stress $\overline{\overline{\sigma}}$ is decomposed into an effective stress $\overline{\overline{\sigma_{ef}}}$ which acts in the preform skeleton and a resin hydrostatic pressure $p_r$. The saturation level $s$ is equal to 0 in the dry preform and between 0 and 1 for modelling the behaviour of the wet preform. $\overline{\overline{I}}$ is second-order identity tensor.

## 3. Numerical studies of the resin infusion process

Prior to validate the numerical model by some comparisons with experimental approaches, sensitivity studies of important manufacturing parameters in the LRI process must be carried out. In order to ensure that basic phenomena can be observed, a plate is considered here. It is a basic geometry classically employed in industry to assess and tune RIP processes.

### 3.1 The basic assumption of resin flow

Even if the thermo-chemical model was proposed in the work of *Celle et al.* [1, 2], as the real infusion processes involve complex mechanical situations on which we focus, isothermal condition was considered here, corresponding to constant resin viscosity. On the other hand, as indicated earlier permeability of the preform is always a key parameter in LCM processes [15, 16], quite tricky to assess even if some recent progress permits to anticipate its introduction in realistic simulations [19]. Here, as a first approximation, the Carman-Kozeny's relation (eq.6) [20] is employed to determine the permeability tensor:

$$\overline{\overline{K}} = \frac{d_f^2}{16\overline{h_K}} \frac{(1-V_f)^3}{V_f^2} \qquad (6)$$

with $d_f$ the average fiber diameter, $\overline{h_K}$ the Kozeny's constant (a vector) and $V_f$ the fiber volume fraction of the preform. It must be noticed that this permeability will change in our simulations, since the fiber volume fraction is updated with respect to the preform deformation all along the process. It is one of the great advantages of setting a general 3D framework to couple resin flow with preform deformation [1, 2, 21].

**3.2 Boundary conditions**

The boundary conditions for simulating the infusion of a plate by LRI process are shown in Fig. 3. For the solid system, at the beginning of the infusion, the vacuum bag creates a mechanical boundary pressure on the surface of the preforms (Fig. 3a). Both displacement and stress vector continuity are prescribed between the flow enhancement fabric and the preform. Zero in-plane displacements of the preforms are prescribed on lateral edges. Applying the vacuum, strong deformation through the thickness of the preform is observed (see Fig. 3b, remeshing in the zone of the prefrom and considering no deformation of flow enhancement fabric) and the resin enters the flow enhancement fabric. Resin normal velocity and pressure boundary continuity are enforced on the interface between the pure resin area (zone of the Stokes flow) and the wet perform (zone of the Darcy flow). Resin velocity is null on both sides of the preform due to the presence of the vacuum bag.

Figure 3 about here

### 3.3 Numerical sensitivity studies

Unless specified, a preform with the dimensions of 335 mm × 335 mm × 20 mm was used in these numerical studies. Moreover a constitutive law of the dry preform corresponding to the NC2 (Non Crimp New Concept produced by Hexcel Reinforcements) fabrics mentioned in [1] was employed during the compaction phase. The resin viscosity and the initial porosity of the preform (before compaction) are equal respectively to 0.03 Pa.s corresponding to a RTM6 resin at 120°C, and 60%.

### 3.3.1 Test of convergence

Convergence tests allow us to indicate the required number of elements to be used. In this test, we mainly observe the evolution of filling time versus the number of elements in the structured mesh (Fig.4), as this evolution is usually more important than the other output parameters.

The blue curve on Fig. 4 shows that for a number of elements larger than 900 (the number is always computed after remeshing), the filling time is stabilized in the numerical simulations. This necessary test was performed before every numerical simulation. The other two lines on this figure were obtained by an analytical approach corresponding to a constant thickness $h$ and a constant isotropic permeability of the preform $K$ :

$$t = \frac{h^2}{2KP}\eta\phi \qquad (7)$$

where $t$ the resin filling time, $h$ the thickness of the preform (a constant), $K$ the permeability of the preform (a constant) and $P$ the pressure differential between the resin inlet and outlet located at a distance $h$ from each other. This analytical expression can be deduced straightly from the Darcy's law (eq.3) assuming constant properties and a constant pressure gradient. Since the analytical approach can not take into account the variation of the thickness of the prefrom during the filling phase, the maximum (after infusion) and minimum (after compaction) thicknesses were chosen for the calculations (see the analytical results 1 and 2 on figure 4) and then compared with the numerical simulations results. Moreover the average permeability obtained by the numerical simulation ($3.29 \times 10^{-14}$ m$^2$) was employed in these analytical approximations. Finally, the filling time in the stable zone of the numerical simulations is well bounded by analytical results, due to the evolution of the thickness of the preform during the resin infusion stage.

Figure 4 about here

### 3.3.2 Changes in the geometric dimensions

*For the draining fabric*

The results of numerical simulations based on the change in initial thickness of the draining fabric are shown in Table 1. The dimensions of the preform remain constant: 335 mm × 335 mm ×20 mm. On the contrary, another constitutive law of the preform in compression is employed, corresponding to the material used in the experiments (see §4.1 below). The corresponding permeability after compaction stage is $5 \times 10^{-14}$ m$^2$. We note that changes in the thickness of the draining fabric almost do not disturb the numerical results. It yields a small variation of the filling time (2%), which corresponds mainly to the evolution of the mesh density of the structure. From an experimental point of view, it is assumed that the thickness of the draining fabric can not change during the infusion process, but this will not affect the manufactured plate anyway.

Table 1 about here

*For the preform*

In our numerical model, the thickness of the preform is more important than any other geometrical parameter. Under industrial conditions, changes in the thickness of the preform generate a series of variation of major process parameters after the infusion stage. Table 2 shows the numerical simulation results corresponding to the different thicknesses of the preform. As a different constitutive law in compression was used in this numerical simulation, a plate thicker than the one computed in the previous tests was obtained after the filling stage. We notice that when the initial thickness of the preform varies, filling time, resin mass absorbed and final thickness of the preform change unlike the fiber volume fraction that depends solely on the preform behavior and the initial porosity. As expected, if the thickness increases, it requires longer time and more resin to infuse completely the preform. The evolution of the filling time versus the changes in initial thickness of the preform is shown in Fig. 5. A non-linear evolution was obtained as expected from the simple relation (Eq.7). Complementary studies permitted to verify that varying the length and width of the plate affected only the mass of resin absorbed.

Table 2 about here

Figure 5 about here

## 4. Experimental Approach

For the experiments, a RTM6 resin was considered together with G1157 "unidirectional fabrics" reference G1157 produced by Hexcel Corporation.

### 4.1 Characterisation of the dry preform response

To characterize the dry preform behaviour before resin infusion, an independent test of transverse compression with the G1157 UD used in the following LRI test (48 plies composite plates $[0_6\ 90_6\ 90_6\ 0_6]_s$) was achieved in the laboratory of Hexcel Corp. on a Zwick Z300 (300 kN) machine. The experimental curves of force versus displacement through the thickness of the preform were obtained. Then Cauchy stress in the fabrics normal direction was expressed as function of corresponding logarithmic strains such as presented in Fig.6. The compression results show that dry fabrics have a strongly non-linear behaviour.

Figure 6 about here

### 4.2 Plate Liquid Resin Infusion test

Infusion experiments were conducted with 48 plies composite plates $[0_6\ 90_6\ 90_6\ 0_6]_s$, made up of G1157 UD. The dry preform dimensions are 335 mm × 335 mm × 20 mm and the total mass measured is 1.56 kg. The experimental setup used to characterize the infusion test is shown in Figure 7. This infusion test was carried out under standard industrial conditions, using a heating plate with an upper lid to guarantee homogeneous thermal conditions. Before infusion, the resin is preheated to 80°C in a heating chamber, while the preform is heated at 120°C. The resin entry and exit are presented also in this Figure 7, and a balance is used to measure the resin mass absorbed by the whole system during the infusion stage. A micro-thermocouple (TC1) is inserted in the middle of the entry tube to monitor the resin temperature and detect the initial filling time. To initiate the measurement of the resin mass, another micro-thermocouple (TC2) associated with the mass capture unit is placed at the same location as TC1. In the outlet pipe, micro-thermocouple (TC7) is used to monitor the temperature change of the resin outlet and therefore to determine the filling time. To detect the temperature of the preform and the resin flow front during the filling stage [22, 23], 4 micro-thermocouples (TC3-TC6) are placed across the thickness of the preform, at the center of ply 10, ply 25, ply 40 and ply 46 respectively. The ply number (1 to 48) is defined from the flow enhancement fabric (draining fabric) towards the bottom of the preform (heating plate). All the thermocouples are located at the center of the ply, along the same direction as for the carbon fiber to minimize intrusivity [22, 23].

Figure 7 about here

### 4.2.1 Temperature of resin inlet and outlet during the filling stage

Fig. 8 shows the temperature change measured by micro-thermocouple TC1. From this measure, we can not only identify changes in the temperature of the resin entry but also detect time 0 of the filling stage. As indicated previously, because of the difference in temperature between the resin and the preform, a drop of temperature is observed, which indicates the time of resin arrival in the entry tube at 52 s corresponding to time 0 of the temperature measurement and resin flow front detection. After this time, resin inlet temperature increases due to the effect of the heating plate, but it remains at a fairly low level, between 82°C and 92°C.

Figure 8 about here

As indicated previously, micro-thermocouples TC2 and TC7 are placed in the entry and exit tubes respectively. They can monitor both inlet and outlet temperatures while the mass of the resin is measured. Fig. 9 gives the temperature changes of these two thermocouples versus the filling time. From them, it is possible to obtain time 0 of the resin mass measurement (at 70 s) and to estimate the duration of the infusion stage (1100 s). Temperature evolution of TC2 is identical to that of TC1 since they are fixed at the same place. Regarding the temperature of the resin outlet (measured by TC7), we obtain a rather stable evolution, between 104°C and 108°C. The increase in temperature after 1170 s indicates when the resin exits from the preform and enters the outlet tube, as the resin is warmer than the empty tube. The end of the filling stage can be deduced and the total filling time is about 1100 s.

From an experimental point view, a standard LRI process should be performed under a closed lid and in an oven [25] in order to obtain an isothermal condition. After a combination of the resin temperature curves in this close-lid test (Figure 9, resin inlet temperature: 83°C-88°C and resin outlet temperature: 104°C-108°C) and another LRI test in an oven (resin temperature : 99°C-103°C [25]) and a comparison of the dynamic viscosity evolution of the RTM6, a resin viscosity of 0.058 Pa.s corresponding to 100°C (given by Hexcel Corporation) was employed in the following numerical simulations.

Figure 9 about here

### 4.2.2 Resin flow front across the thickness of the preform

Temperature evolutions of the preform during the filling stage are given in Fig. 10. Time 0 corresponds to the initial point of the measurement (deduced from TC1 placed in the inlet tube, see Fig. 8). Before resin infusion, a temperature gradient of about 0.3°C/ply is found across the thickness of the preform. The temperatures tend to decrease when the test begins and the 'cold' resin is left free to fill in the preform. These temperature signals decrease more and more when the resin front flows gets closer to the thermocouples, until the minimum temperature is obtained when resin flows over the thermocouples. The times when the minimum temperature is reached for each thermocouple are indicated in Figure 10. It reveals the resin flow front positions in the preform as demonstrated in [22, 23]. As the heating plate continues to heat up the whole infusion system, temperatures increase again. Specifically, one can note for TC5 and TC6, placed at ply 40 and 46, a small zone where the temperature increases rapidly at about 940 s. This may be related to the heat conduction effects due to the contact between the heating plate and the resin.

From these information the change in resin flow front position across the thickness will be figured out and compared with the one calculated by numerical simulation (see section 5.3).

Figure 10 about here

**4.2.3 Resin mass absorbed during the filling stage**

The change in resin mass versus the filling time during the infusion stage is shown in Fig. 11. One can verify out that the infusion rate decreases during the filling phase for a constant pressure differential applied. At the beginning, the resin enters quickly the tube and then the draining fabric; it corresponds to the strong slope in the first part of the curve. Regarding the whole filling duration estimated from TC2 and TC7 (see Fig. 9), the mass absorbed by the resin infusion system is 705 g. With the industrial requirement to ensure a complete infusion, the inlet tube was not closed immediately after the resin entered the outlet pipe at the end of the filling stage. Consequently, the resin mass still increases after 1100 s. This additional infusion phase takes usually a few minutes.

Figure 11 about here

**4.3 Curing and cooling phases**

After the infusion stage, the temperature of the preform is about 120 °C. It is then increased to 180 °C and maintained for about two hours for the curing stage. Finally, the plate is cooled down to room temperature. A measurement of the average thickness of the final plate (measured in 25 points) shows 12.11 mm with a coefficient of variation of 6.36%. The fiber volume fraction of this plate is then estimated to 62%. Further estimates show that the void defects in the final plate are about 0.7%. Although several micro-thermocouples are present in the preform, a quite low void content is obtained in our composite part. It can be verified again that the micro-thermocouples used in our experimental studies to monitor the resin infusion process have a negligible intrusivity [22, 23].

## 5 Comparison of numerical simulation and experimental analysis

**5.1 Input simulation parameters**

Concerning the geometrical parameters of the preform, they were already presented previously: the initial thickness of the preform was measured at 20 mm before the compression under the vacuum bag. The surface dimensions are 335 mm × 335 mm. The initial fiber volume fraction of the preform was calculated at 39%. The constitutive response under compression corresponding to fabrics G1157 was determined in section 4.1 (see figure 6). Regarding figure 9, a resin viscosity of 0.058 Pa.s corresponding to 100 °C was chosen (see section 4.2.1)

In order to realize a representative numerical simulation of the real infusion test, an analytical expression of the saturated transverse permeability ($K_s$) of the unidirectional fabric G1157 used in the infusion test is introduced; it was obtained experimentally by *Nunez* [24] as a function of fiber volume fraction $V_f$ knowing both total thickness and areal weight (eq. 8).

$$K_s = 1.626 \cdot 10^{-11} \cdot (V_f)^3 - 2.815 \cdot 10^{-11} \cdot (V_f)^2 + 1.474 \cdot 10^{-11} \cdot V_f - 2.048 \cdot 10^{-12} \qquad (8)$$

**5.2 Simulation results and comparisons with experimental data**

Simulations have been realized with 1458 triangle mixed velocity-pressure elements. Adequate boundary conditions were used to represent, as properly as possible, the industrial environment (see Fig.3). Experimentally, resin infusion has been performed under a vacuum pressure of 1.4 mbar.

The numerical results and a comparison between the experimental and numerical simulation are given in Table 3. Generally, a good agreement can be observed between these two studies for the major parameters. Since we achieved a standard plate infusion test with closed lid, the thickness variation of the preform could not be assessed unlike in previous studies [23]. A second comparison was realized in an open-lid infusion test carried out on 24 plies fabric G1157 composite plates and is presented later.

Changes in resin mass used during the filling stage was detected by the mass acquisition unit (see Fig. 7), even if this is a mass absorbed by the whole infusion system. It can be considered that little resin remains in the draining fabric after a complete infusion stage. Regarding the filling time, both experimental and simulation results are close (13% of difference). It must be noticed that the filling time depends strongly on both permeability and resin viscosity, two major input parameters quite tricky to assess. For the experimental value in Table 3, the filling duration of the preform is estimated by removing the time required for the filling of the draining fabric (100 seconds) at the beginning of the infusion test.

Table 3 about here

Another comparison of a plate infusion test with 24 G1157 plies is presented in Table 4. The experimental protocol and properties are almost the same as the test mentioned above. This test yields two additional information: (1) numerical simulation was performed for the resin infusion test with a different thickness of the preform, (2) variation of the thickness of the preform during the filling stage could be measured by a fringe pattern projection technique under different experimental conditions (lid open) [23]. Similarly to the previous comparison, the numerical simulation and experimental analysis are very similar in the compaction phase. For the evolution of the thickness and the fiber volume fraction, a good correlation can be noted between the numerical and experimental approaches. The resin mass is directly related to the thickness of the preform, consequently it also leads to a satisfactory correlation with that of numerical simulation. However the difference in filling time is more pronounced here (18%). Indeed, it is well known that poor thermal conditions lead to resin viscosity increase, and longer infusion stage.

Table 4 about here

**5.3 Resin flow front evolution during the filling stage**

Fig. 12 presents the experimental, numerical and analytical results concerning the position of the resin front during the filling stage of the 48 plies plate infusion test (see section 4.2). For the numerical simulation results, to asses the flow front position 5 nodes were selected across the stacking thickness of the preform, on a same line corresponding to the TCs position (at the center of ply plane, see section 4.2). On the other hand, experimentally 4 micro-thermocouples (TC3-TC6) have been used to characterize the resin flow by measuring changes in temperature of the preform [22] (see Fig. 10). Here we should point out that the resin takes about 50 seconds to arrive in the middle of draining fabric, consequently it must be subtracted this time to each thermocouple to determine the evolution of the resin front position through the thickness of the preform.

In figure 12, time of resin arrival at position 100% corresponds to the time when resin is in contact with the heating plate at the bottom of infusion system. In the case of the comparison with analytical results (eq.7), accounting for the thickness variation is mandatory. Considering an average thickness of the preforms may be used as a first approximation, in this very basic geometry. Numerical simulations account for the preform deformation, the average thickness of the preform can be computed (between the maximum and minimum thickness mentioned previously) and be integrated in the analytical calculations. Comparing the three curves of interest, a very close correlation can be noted here for the estimates of resin front position in the middle of the preform between the experimental approach and numerical simulation.

Figure 12 about here

## 6 Discussions

Different manufacturing conditions generate different resin flow and process properties. The effects of varying some important production parameters are discussed here.

**6.1 Variation of the thickness of the preform**

Results of two plate infusion tests (24 and 48 G1157 plies) under standard industrial conditions (with closed lid) were presented which differ by the initial thickness of the preform. Comparisons of the major parameters of the resin infusion process obtained experimentally in these two cases are presented in Table 5. We got generally the final composite plates with almost the same fiber volume fraction, and the other important output parameters of the test with 48 plies are two times those for the test with 24 plies.

It should be pointed out here that the temperature of heating plate in the test with 24 plies ($\approx 115°$ C) is lower than in the test with 48 plies ($\approx 125$ ° C). We can postulate straightly that in the test with 48 plies, the higher temperature of heating plate shortens the duration of the filling stage by lowering the resin viscosity. Normally, the filling time is not proportional to the preform thickness. It is confirmed by the previous numerical simulation results (see Fig. 5). Compared with the test with 48 plies, there was only half of resin mass absorbed during the infusion stage with 24 plies, as this parameter depends strongly on the porosity and the volume of the preform. On the contrary, for the thickness of the final plate, there are not such relations between these two infusion tests. Thickness is also related to several other parameters, for example, the vacuum level, the resin mass, the curing rate and so on.

Table 5 about here

**6.2 Change in the resin temperature during infusion**

Temperature of resin flow is another significant parameter in the infusion tests, as the resin viscosity depends strongly on it. The key data of the process have also been compared between two plate infusion tests with 48 plies fabric G1157 under different experimental environments: the close-lid test and the open-lid test (Table 6). Temperature of the heating plate differs very little in these tests. On the contrary, in the open-lid test the low temperature of the resin inlet and high temperature gradient across the thickness of the preform disrupted the resin flow both in the draining fabric and the preform. As the resin could not flow properly during the filling stage, much longer time was necessary to infuse completely the preform and the final composite part presented more porosity. Finally, one can observe a thicker composite plate manufactured with lower fiber volume fraction and that the thickness of the final part is more homogenous in the close-lid infusion test.

Table 6 about here

**6.3 Estimation of the permeability of the preform**

As an essential parameter, permeability of the preform plays an important role not only in the real resin infusion process but also in our numerical model. Normally, we have three possibilities to estimate the permeability of the preform: the simplest way is to postulate a constant permeability during the whole filling duration; another classical method relies on the Carman-Kozeny's equation presented previously (eq.6) that is largely used in the LCM modeling, but the Kozeny's constant should be determined in advance; eventually a more precise way requires an experimental approach to assess the real permeability of the preform, but this measurement often faces difficult experimental problems.

Fig. 13 presents the change in the resin front position versus time calculated by numerical simulation using 3 different methods to determine the transverse permeability of the preform. These calculations are based on the one mentioned previously (see section 5) corresponding to a standard LRI process with 48 G1157 plies with closed lid presented in the section 4.2. A nice correlation is noted for the numerical simulation results obtained through the Carman-Kozeny's equation (eq.6 and $\overline{h_K}$ = 10) and the experimental analytical expression (eq.8) respectively. The Carman-Kozeny's equation can be adopted in the numerical analysis for our resin infusion test cases.

On another hand, an important difference in the resin front evolution could be observed when we compared the results obtained with a constant permeability ($4\times10^{-14}$ $m^2$) corresponding to permeabilities calculated from the average porosity through eq.8 and the experimental expression (eq.8). At the beginning of the infusion stage, it can be observed obviously that the resin flows more rapidly in the case of a constant permeability ($4\times10^{-14}$ $m^2$), since the permeability obtained experimentally presents a lower value. On the contrary, the analytical permeability deduced from the experiments becomes greater than $4\times10^{-14}$ $m^2$ in the last part of the infusion stage. As a conclusion, a pure estimate of the permeability will change directly the filling times. Here filling times are 27% longer for constant permeability. This highlights also the need for simulations accounting for preform deformation, and hence permeability update during the infusion stage.

Figure 13 about here

# 7 Conclusions

In this paper, numerical and experimental analyses of the resin infusion manufacturing process were presented. Subsequently, some general comparisons between numerical simulation and experimental results were realized based on a plate infusion test by LRI process under industrial conditions. From these comparisons and some additional discussions on the most important process parameters, we consider that our numerical model is able to deal with the problem of interaction between resin flow and the deformations of the porous performs during the resin infusion stage. It has also been demonstrated that only a numerical model is able to handle preform compaction during fluid infusion and is able to account for permeability variation and hence yield realistic filling times.

Although a good correlation can be obtained between numerical simulation and experimental approach, some problems remain to be solved both in the numerical computations and experimental measurements. One main problem corresponds to characterising the industrial conditions, such as the resin viscosity, the preform thickness before compaction, and more generally the thermal environment. The next step in this validation process will hence focus on thermal and chemical aspects of LRI processes.

**Acknowledgements**

The authors would like to thank HEXCEL Corporation and ESI group for their support to the present research work.

List of figures

Fig. 1. Principle of Liquid Resin Infusion process

Fig. 2. Different zones making up a numerical model of the resin infusion processes: (a) RFI, (b) LRI

Fig. 3. Boundary conditions used in numerical simulations of a LRI plate infusion test (a) for the solid system ($\vec{u}$: displacement) and (b) for the resin flow after the compaction phase ($\vec{v}$: fluid velocity)

Fig. 4. Evolution of the filling time vs. the number of elements in the structured mesh

Fig. 5. Changes in the filling time vs. the initial thickness of the preform

Fig. 6. Compression curve in out-of plane direction for dry UD fabric G1157

Fig. 7. Experimental set-up for characterisation of plate infusion test

Fig. 8. Temperature evolution of the resin inlet measured by the micro-thermocouple TC1 during the filling stage

Fig. 9. Temperature evolutions of the resin inlet and outlet measured by the micro-thermocouple TC7 during the filling stage

Fig. 10. Change in time of the signal of the thermocouples placed across the preform

Fig. 11. Resin mass absorbed by the system infusion during the filling stage

Fig. 12. Resin front position vs. the filling time for a standard plate infusion test with closed lid with 48 G1157 plies carried out by LRI process

Fig. 13. Resin front position vs. the filling time computed using 3 different methods to determine the transverse permeability of the preform corresponding to a standard resin infusion process with 48 G1157 plies.

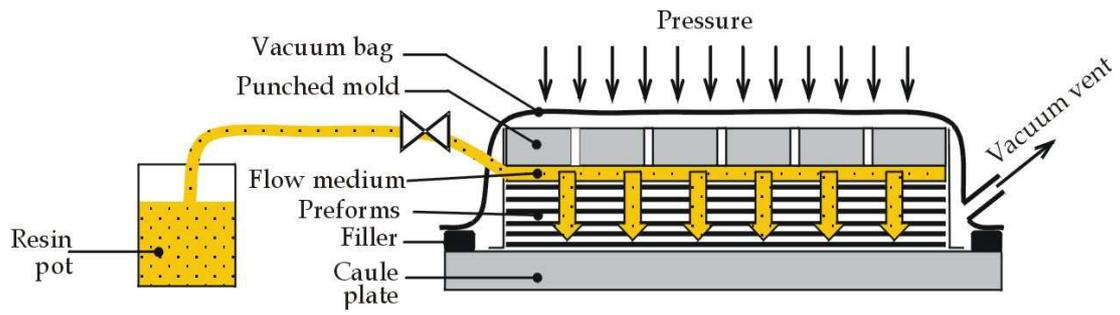

Fig. 1. Principle of Liquid Resin Infusion process

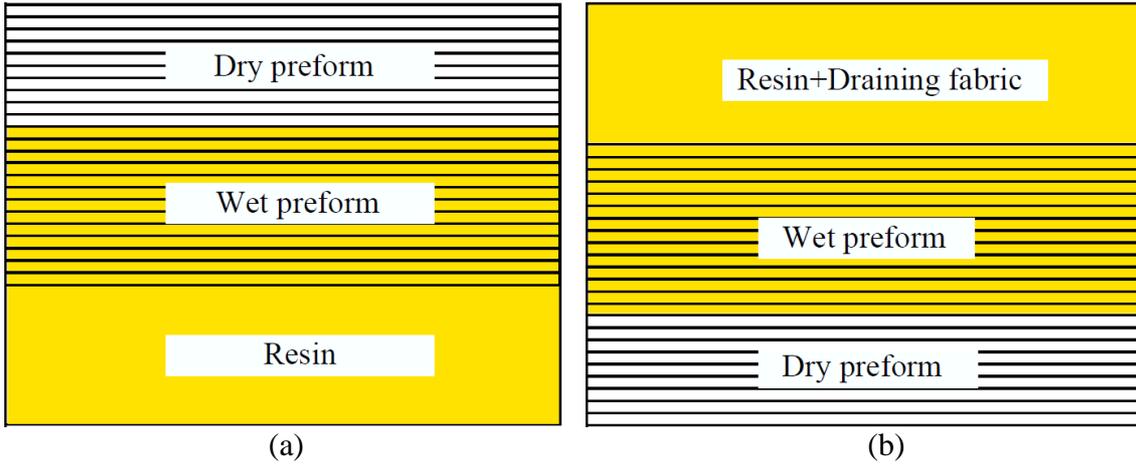

Fig. 2. Different zones making up a numerical model of the resin infusion processes:
(a) RFI, (b) LRI

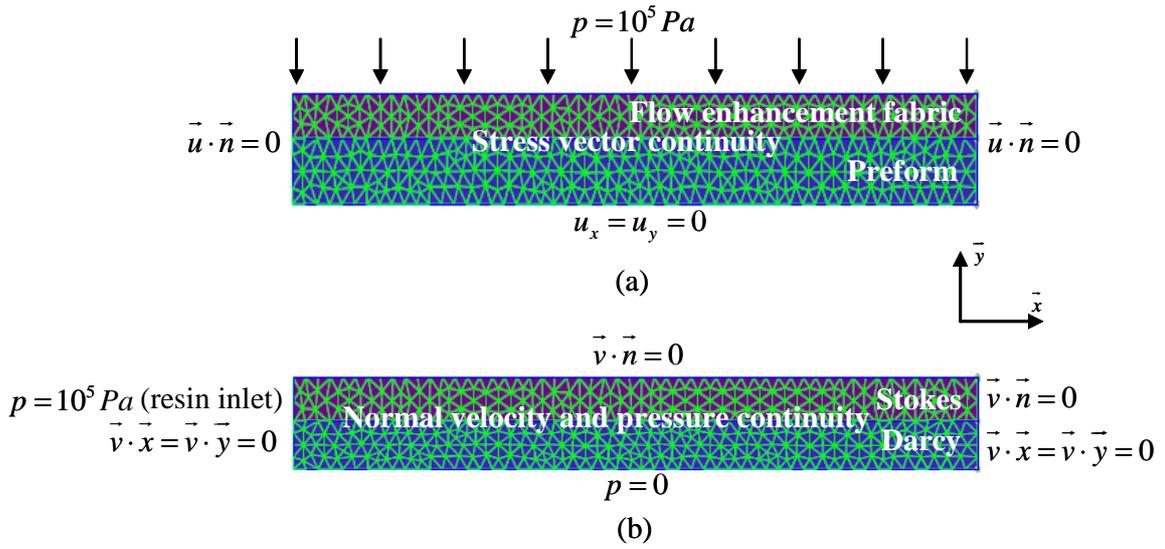

Fig. 3. Boundary conditions used in numerical simulations of a LRI plate infusion test (a) for the solid system ($\vec{u}$: displacement) and (b) for the resin flow after the compaction phase ($\vec{v}$: fluid velocity)

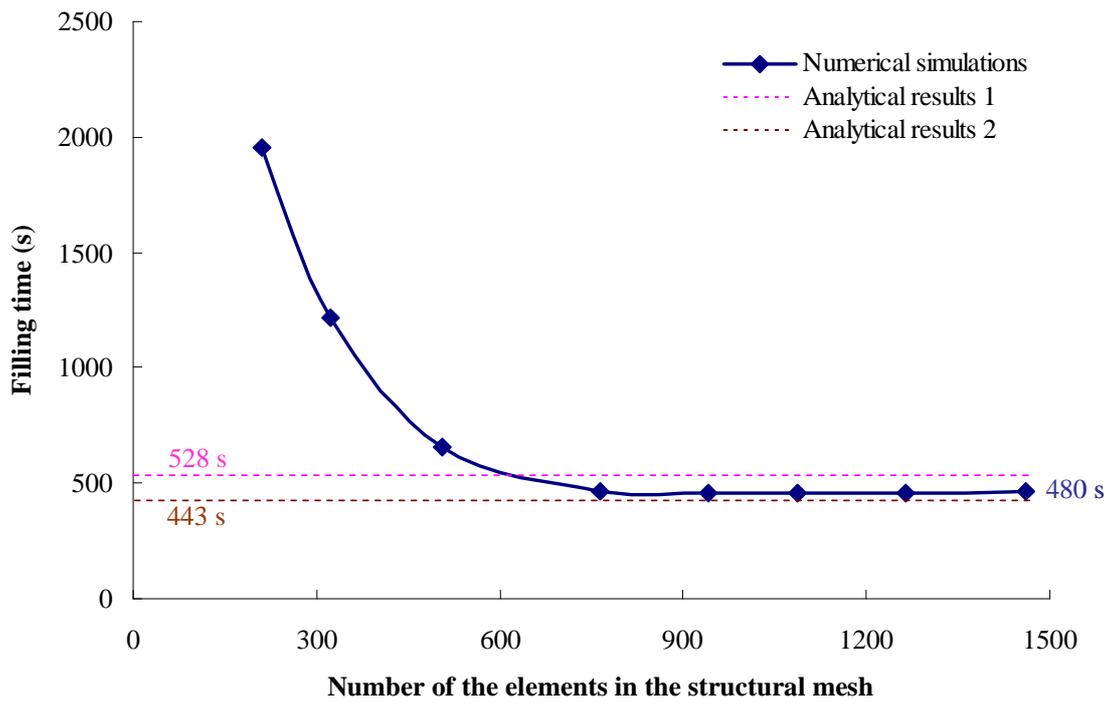

Fig. 4. Change in the filling time vs. the number of elements in the structured mesh

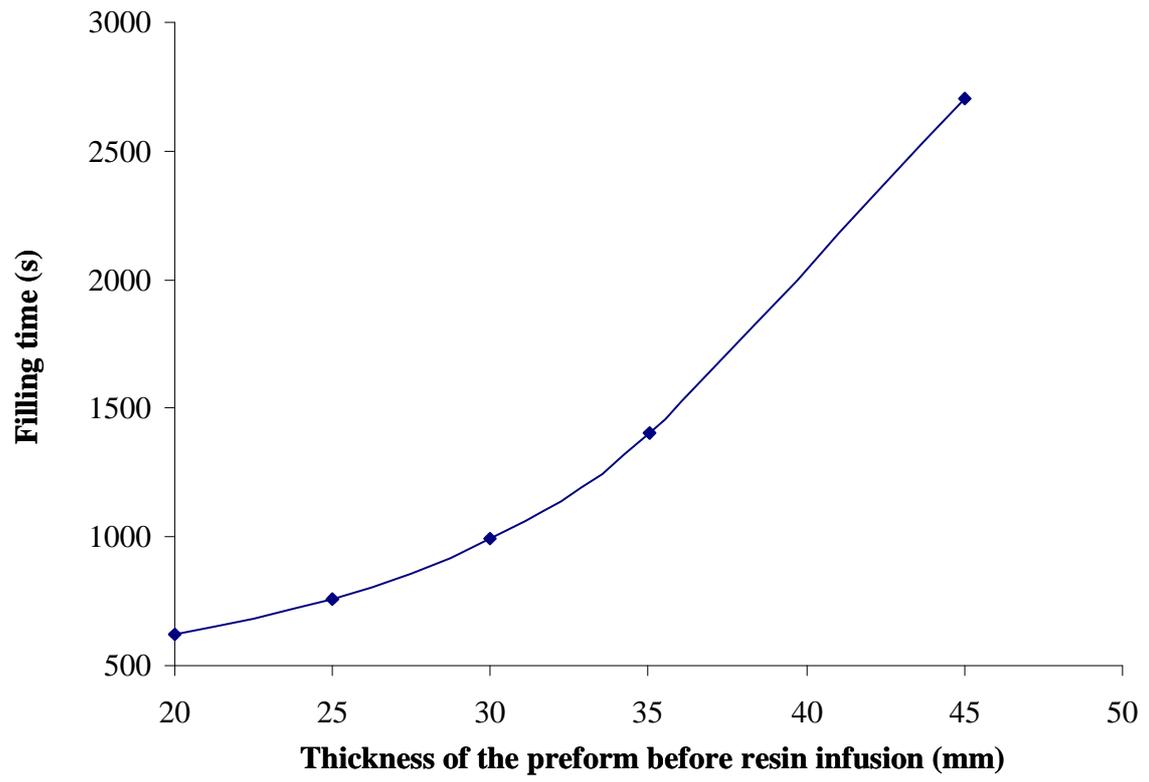

Fig. 5. Changes in the filling time vs. the initial thickness of the preform

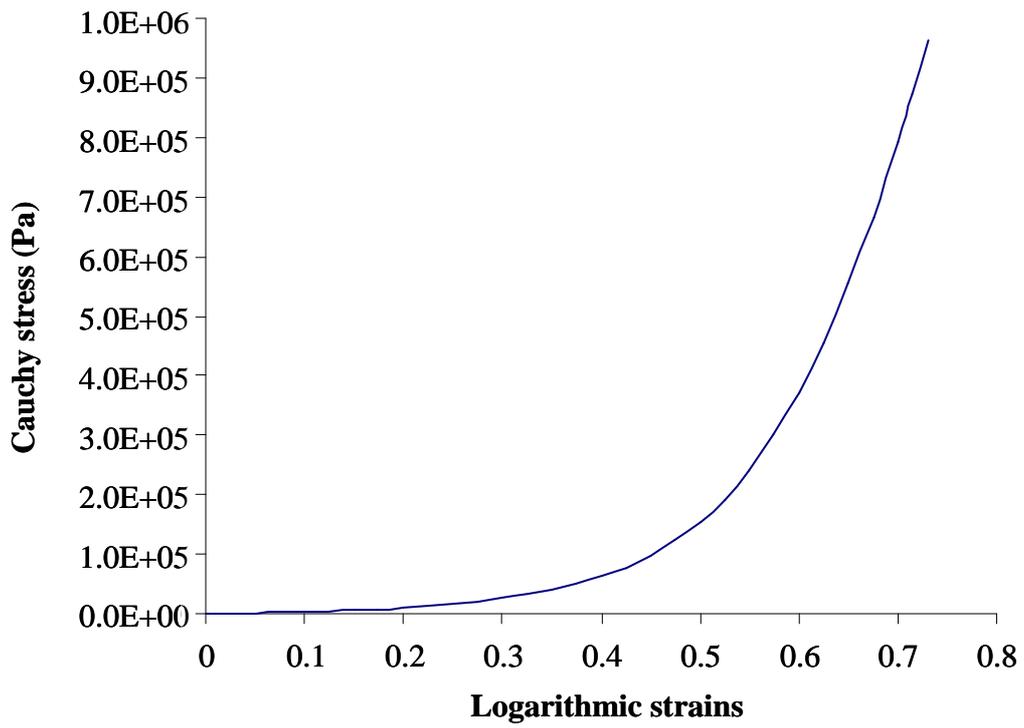

Fig. 6. Compression curve in out-of plane direction for dry UD fabric G1157

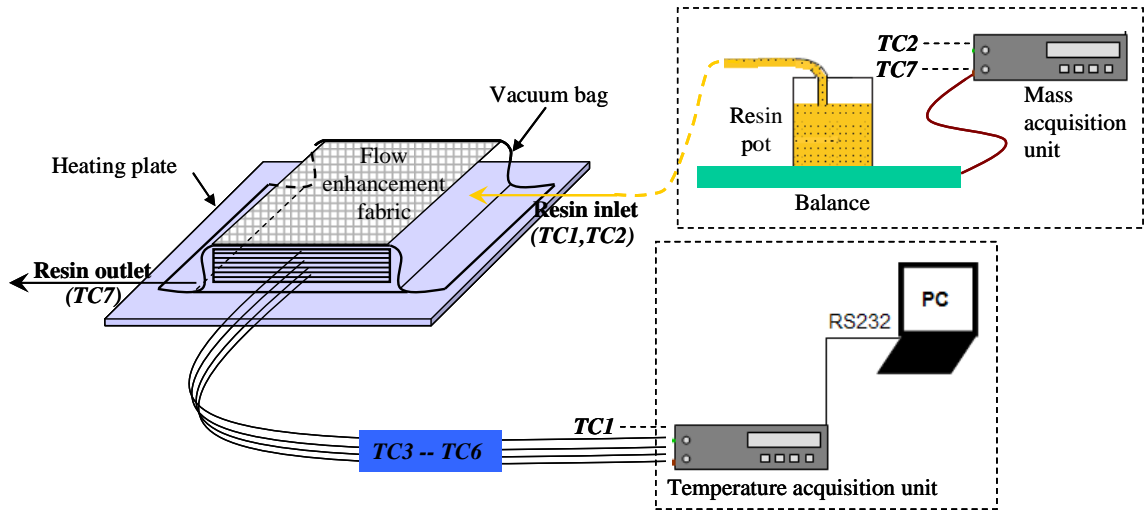

Fig. 7. Experimental set-up for characterisation of plate infusion test

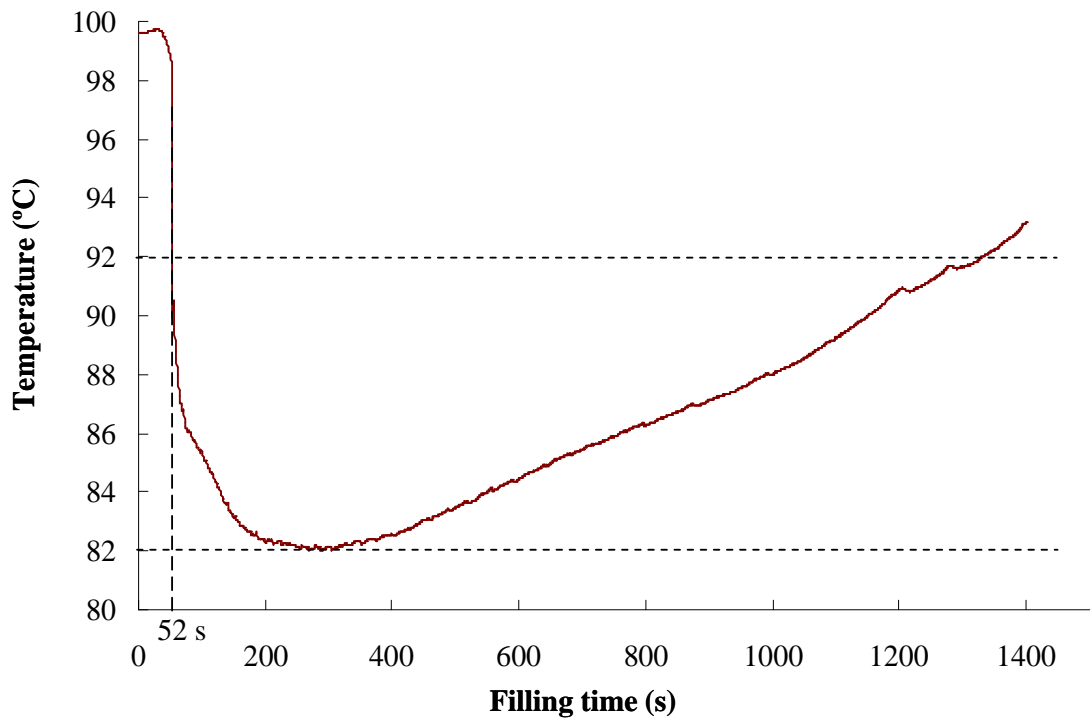

Fig. 8. Temperature evolution of the resin inlet measured by the micro-thermocouple TC1 during the filling stage

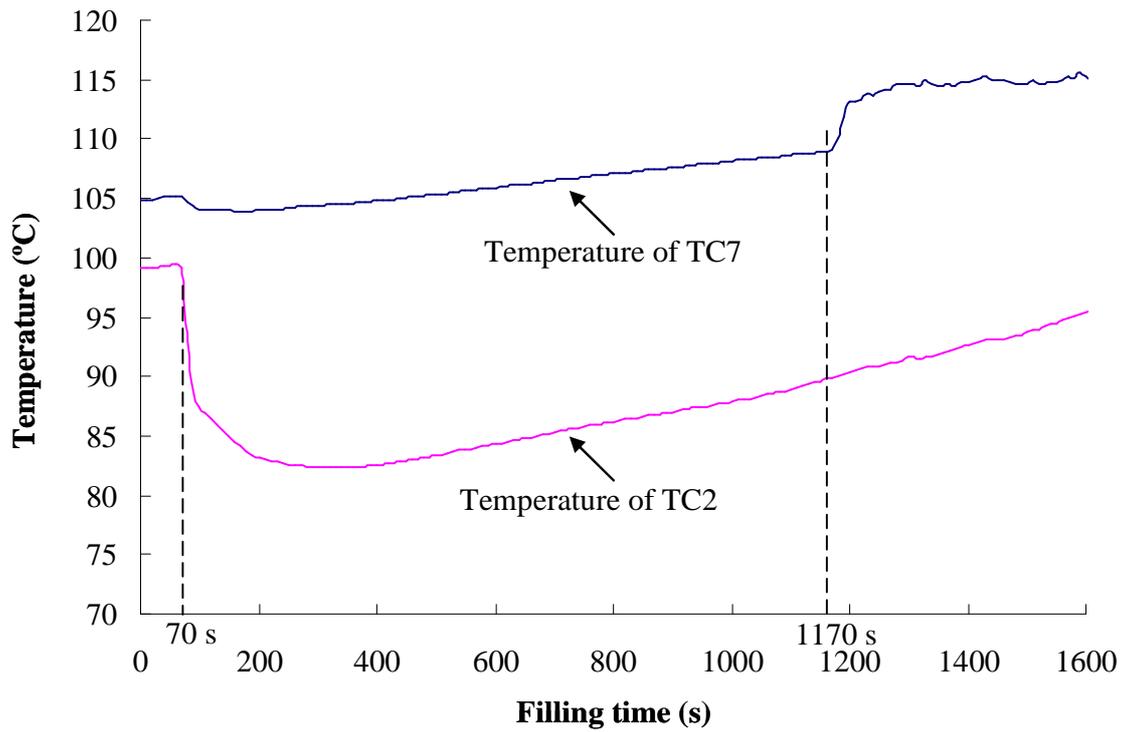

Fig. 9. Temperature changes of the resin inlet and outlet measured by the micro-thermocouple TC7 during the filling stage

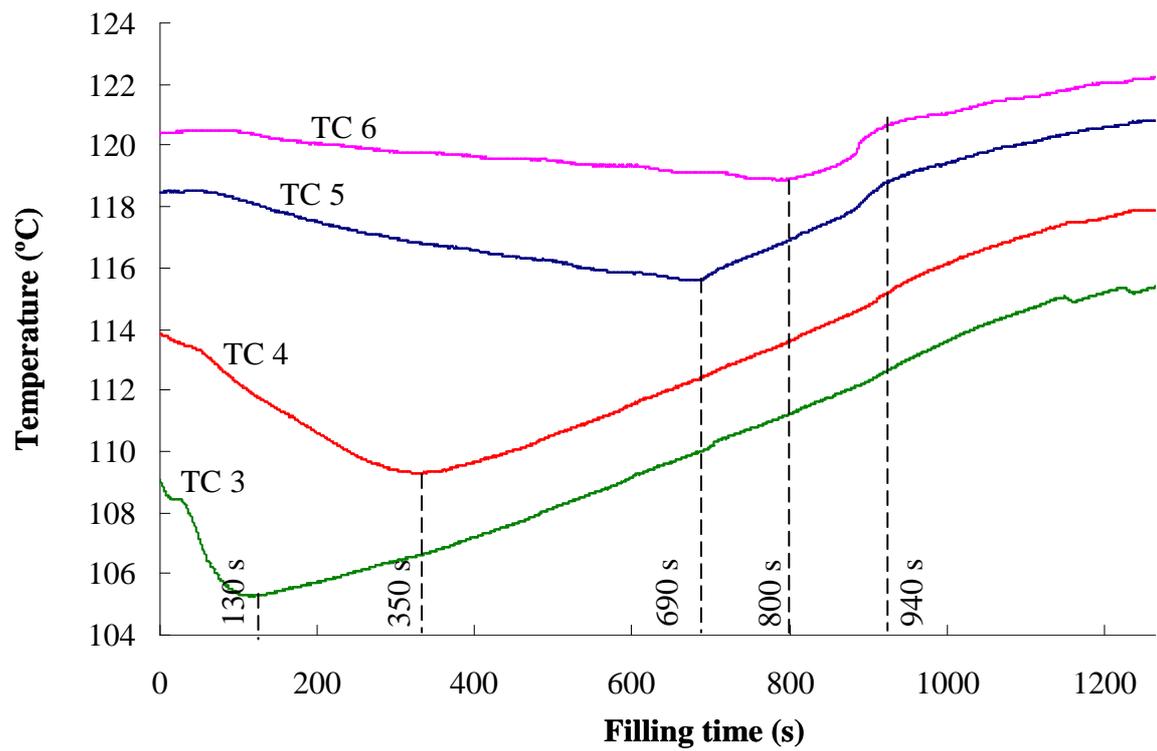

Fig. 10. Change in time of the signal of the thermocouples placed across the preform

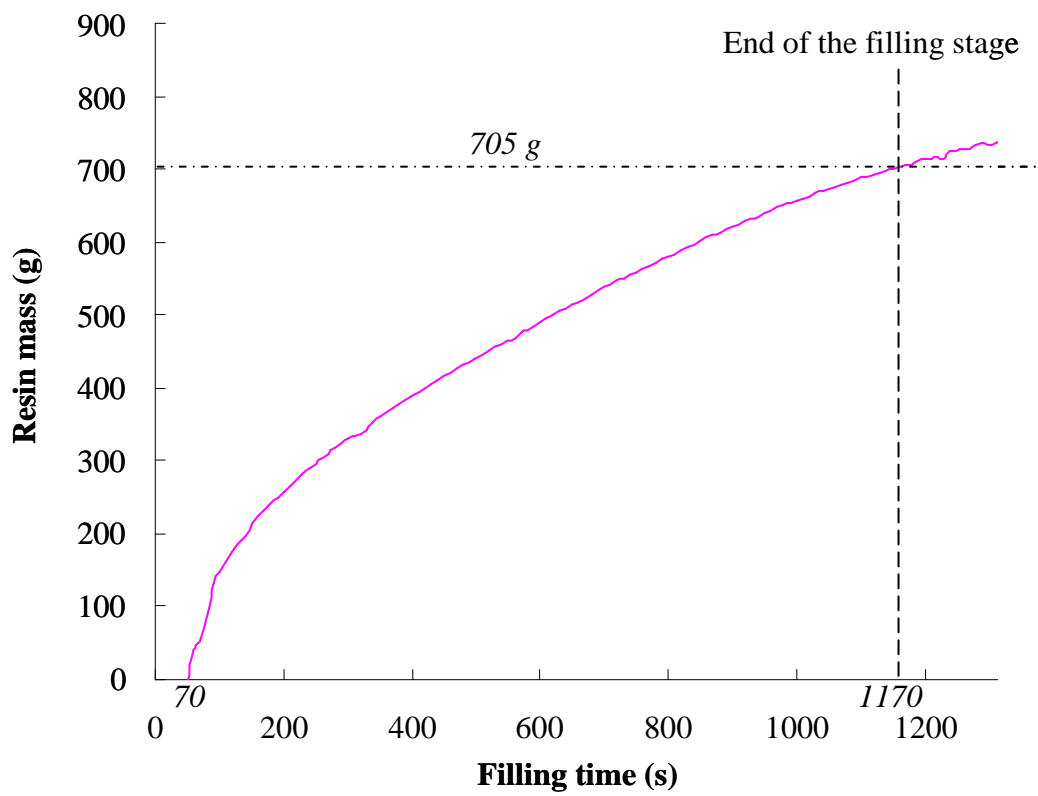

Fig. 11. Resin mass absorbed by the system infusion during the filling stage

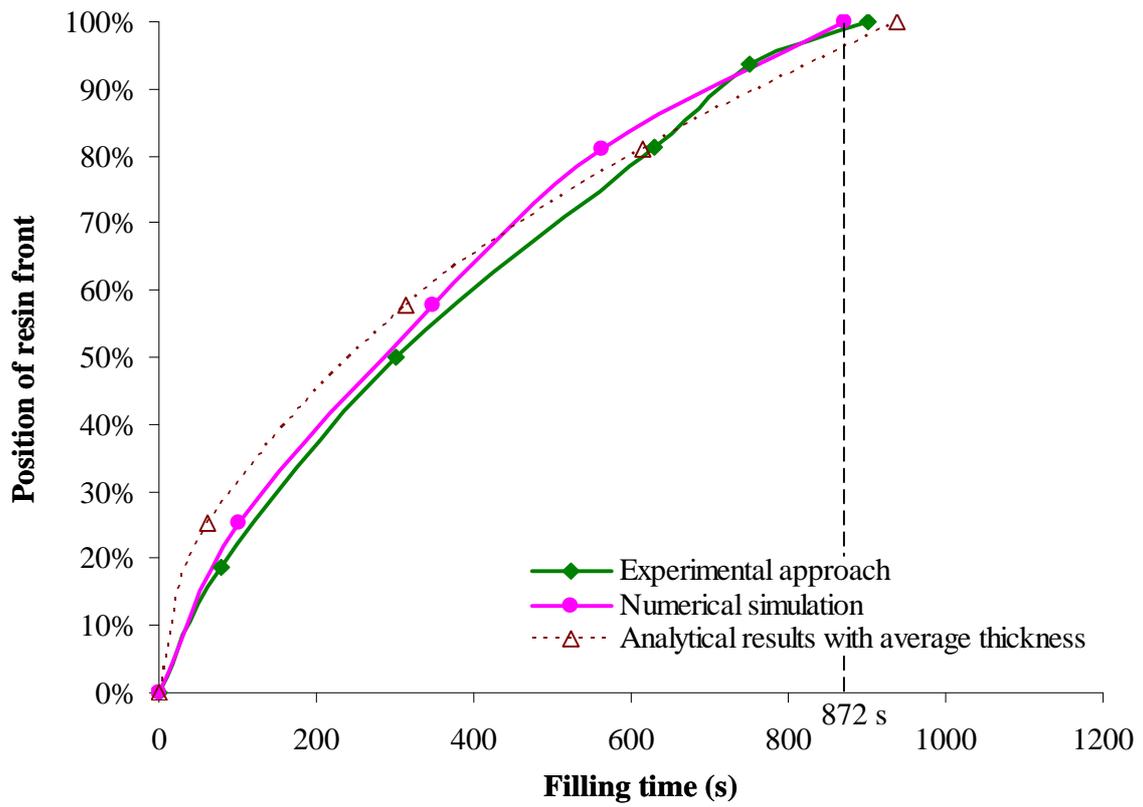

Fig. 12. Resin front position vs. the filling time for a standard plate infusion test with closed lid with 48 G1157 plies carried out by LRI process

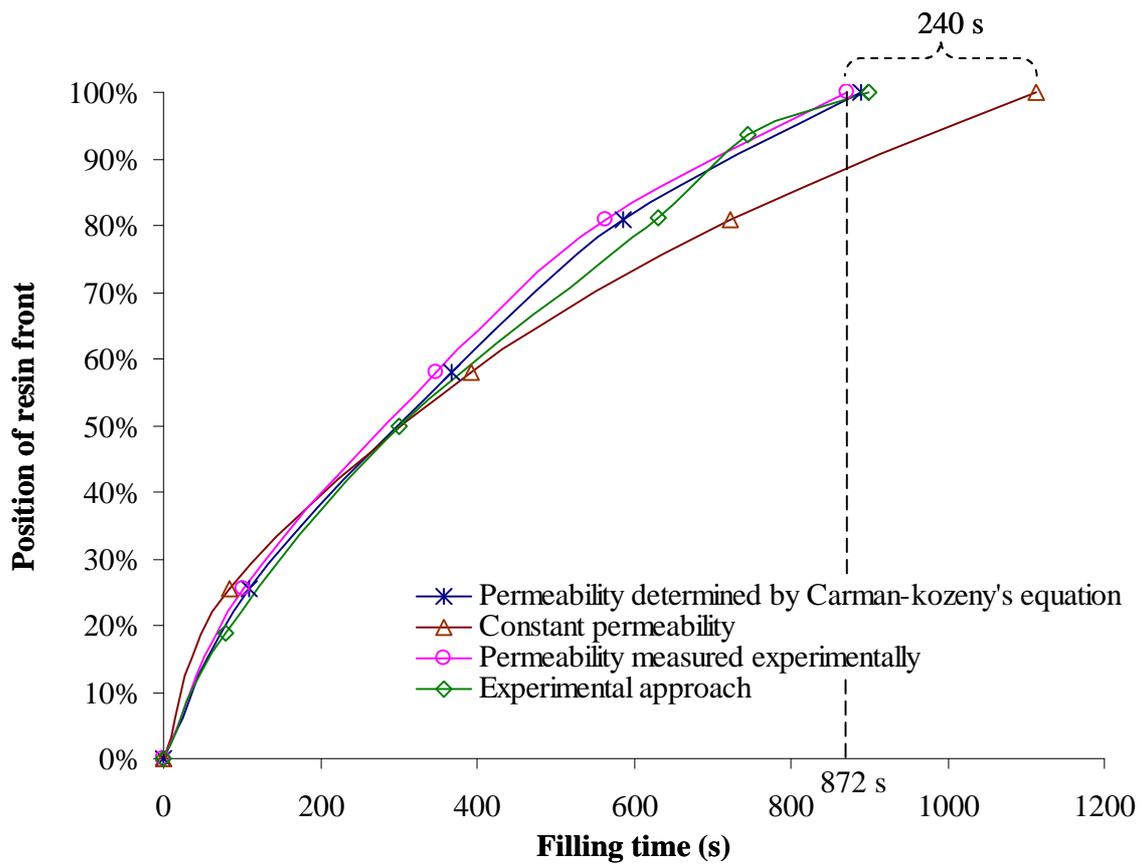

Fig. 13. Resin front position vs. the filling time computed using 3 different methods to determine the transverse permeability of the preform corresponding to a standard resin infusion process with 48 G1157 plies

List of tables

Table 1 Numerical simulation studies according to the variation of the thickness of the draining fabric

| Thickness of the draining fabric (mm) | Filling time (s) | Resin mass absorbed (g) | Thickness of the preform after infusion stage (mm) | Fiber volume fraction after infusion stage |
|---|---|---|---|---|
| 6 | 635.2 | 1245 | 18 | 44.4% |
| 8 | 638.2 | 1245 | 18 | 44.4% |
| 10 | 625.2 | 1245 | 18 | 44.5% |

Table 2 Numerical simulation studies according to the variation of the thickness of the preform

| Thickness of the preform (mm) | Filling time (s) | Resin mass absorbed (g) | Thickness of the preform after infusion stage (mm) | Fiber volume fraction after infusion stage |
|---|---|---|---|---|
| 20 | 625.2 | 1245 | 18 | 44.4% |
| 25 | 758.1 | 1557 | 22.5 | 44.4% |
| 30 | 994.2 | 1868 | 27.0 | 44.4% |
| 35 | 1403 | 2181 | 31.5 | 44.4% |
| 45 | 2703 | 2806 | 40.5 | 44.4% |

Table 3 Comparison between numerical simulation and experimental results of a plate infusion test with 48 G1157 plies carried out by LRI process

| | | Experiment | |
|---|---|---|---|
| Initial condition | Average thickness of the preform (mm) | 20 | |
| | Surface dimension | 335 mm × 335 mm | |
| | Fiber volume fraction | 39% | |
| | Mass of the preform (g) | 1560 | |
| | | Experiment | Simulation |
| After compaction | Average thickness of the preform (mm) | 13 | 12.7 |
| | Fiber volume fraction | 60% | 61.5% |
| After filling | Average thickness variation of the preform (mm) | ／ | 1.25 |
| | Fiber volume fraction | ／ | 56.0% |
| | Mass of resin used during the infusion stage (g) | 705 | 750 |
| | Filling time of preform (s) | 1000 | 872 |

Table 4 Comparison between numerical simulation and experimental results of a plate infusion test with 24 G1157 plies carried out by LRI process

|  |  | Experiment | |
|---|---|---|---|
| Initial condition | Average thickness of the preform (mm) | 10 | |
|  | Surface dimension | 335 mm × 335 mm | |
|  | Fiber volume fraction | 39% | |
|  | Mass of the preform (g) | 780 | |
|  |  | Experiment | Simulation |
| After compaction | Average thickness of the preform (mm) | 6.5 | 6.35 |
|  | Fiber volume fraction | 65% | 61.5% |
| After filling | Average thickness variation of the preform (mm) | 0.55 | 0.6 |
|  | Fiber volume fraction | 55.5% | 56.1% |
|  | Mass of resin used during the infusion stage (g) | 350 | 375 |
|  | Filling time of preform (s) | 500 | 410 |

Table 5 Experimental data of the standard LRI tests with 24 and 48 G1157 plies with closed lid

|  | Infusion test with **24 plies** | Infusion test with **48 plies** |
|---|---|---|
| Initial average thickness of the preform (mm) | 10 | 20 |
| Initial fiber volume fraction of the preform | 39% | 39% |
| Filling time of the preform (s) | 500 | 1000 |
| Resin mass absorbed (g) | 350 | 705 |
| Average thickness of the final composite plate (mm) | 6.25 | 12.02 |
| Standard variation of the thickness of the final composite plate | 5.4% | 4.5% |
| Fiber volume fraction of the final composite plate | 59.5% | 62.4% |

Table 6 Comparisons of the key process parameters under two different experimental conditions for 48 G1157 plies

|  | **Close-lid** infusion test | **Open-lid** infusion test |
|---|---|---|
| Temperature of the heating plate | ≈125°C | ≈130°C |
| Initial temperature gradient across the thickness of the preform (°C/ply) | 0.3 | 0.75 |
| Temperature of resin inlet | 82°C < T < 92°C | 70°C < T < 78°C |
| Filling time of the preform (s) | 1000 | 2870 |
| Average thickness of the final composite plate (mm) | 12.02 | 12.41 |
| Standard variation of the thickness of the final composite plate | 4.5% | 7.6% |
| Fiber volume fraction of final plate | 62.4% | 59.8% |

# Responses to the reviewers
*(Paper for Journal of Composite Material, Ref. No.: JCM-10-0654)*

The authors wish to thank the reviewers for their comments and corrections that have been thoroughly examined. Below are given the responses to the reviewers' demands and remarks.

Reviewer(s)' Comments to Author:

## Reviewer: 1

Comments to the Author:

This paper deals with an important topic: the analysis of fabric deformation and resin flow in composite manufacturing with flexible tooling. In particular, the experimental measurement may be an important contribution. The paper cannot be published, however, as it is in the current form. The reviewer suggests the paper be accepted for publication only after the following points are addressed.

1. The reference list is incomplete. The authors should include in "Introduction" some important references about the analysis of fabric deformation and resin flow. Some comments on the originality of the current paper (compared with these papers) should be made.

• Hubert, Poursartip, A review of flow and compaction modeling relevant to thermoset matrix laminate processing, Journal of Reinforced Plastics and Composites, (1998), 17, 286-318.
• Loos, Rattazzi, Batra, A three-dimensional model of the resin film infusion process, Journal of Composite Materials, (2002), 36, 1255-1273.
• Li, Tucker III, Modeling and simulation of two-dimensional consolidation for thermoset matrix composites, Composites Part A, (2002), 33, 877-892.

Author's answer: The papers mentioned by reviewer have been added in the reference list [26, 27, 28], some comments were made in the "resin infusion modelling section".

2. Equation (5) (Terzaghi's law) was derived considering the force equilibrium in Z (through-thickness direction). Considering the original force equilibrium equation ($dSij/dj=0$), the stress equilibrium equation in Z should be $Szj/dj=0$ where $j=x,y,z$. Remind that Terzaghi's law was obtained from $dSzz/dz=0$, by ignoring the shear stress gradients ($dSzx/dx=0$ and $dSzy/dy=0$). This may be valid in the case of soil mechanics where the shear stiffness of the porous medium is negligible (e.g. soil bed). However, the shear stiffness of fabric reinforcements may not be negligible in the manufacturing of high performance composites. In most of modeling and simulation works, Equation (5) was assumed, whereas the above three references presented the influence of shear stiffness and shear stress gradient on the force equilibrium. I suggest the authors make some comment on this point.

Author's answer: We do not ignore the shear response of the fabrics. The macroscopic stress in the wet preform can be described, through a 3D extension of the genuine Terzaghi's model which is in fact a mere superposition of hydrostatical components of the fluid and solid phases but can be sufficient in this approach as demonstrated several times by many other authors :

$$\sigma_{ij} = \sigma_{ij}^{ef} - sp_r \delta_{ij} \quad (i,j = 1,2,3)$$

Hence, the full 3D response of the fabric is represented, accounting 'only' for the resin hydrostatic pressure which indeed should be introduced in further refinement of the response.

3. The authors concluded that they obtained good agreements between numerical prediction and experimental measurement. It is hard, however, to accept since Tables 4-5 show great discrepancies between numerical prediction and experimental measurement of mold filling time (1000 vs. 872 in Table 4 and 500 vs. 410 in Table 5). As shown in Figs 9-11, the resin temperature was neither uniform nor constant, whereas the isothermal mold filling simulation was performed. Hence, this might yield some error in the assumption of isothermal condition used in the simulation. In this work, the constant resin viscosity at the temperature 100°C was applied whereas the temperature values measured by thermocouples TC3-TC6 (Fig. 11) were 105-120°C. Some comment should be made on the limit of isothermal mold filling simulation. I understand that the authors developed the thermo-chemical model in the previous works (references [1], [2]) as the authors indicated in Page 7. What prevented the authors from using this model?

Author's answer: We agree with the reviewer's comments: it is hard to compare the numerical simulation and experimental measurement regarding the complexity of the process and the process parameters variability (temperature -> viscosity, permeability, …).

Filling time: Tables 3 and 4 (in revised version) show as whole good agreements between these two techniques – text changed accordingly. Experimental results of filling time, 1000 s in Table 3 and 500 s in Table 4 show the time when resin exits from the prefrom and enters the outlet tube. But numerical simulation results of filling time (872 s in table 3 and 410 s in table 4) indicate the time when resin flow arrives at the bottom of the prefrom, this resin flow is uniform. However, Figure 13 shows a good agreement between numerical and experimental flow front positions across the thickness at the centre of the preform.

Yes, experimental measurements showed that the resin temperature is neither uniform nor constant. We have some important reasons to choose an isothermal filling numerical model:
1. From an experimental point view, we want to realize an isothermal filling stage, so a standard LRI test is carried out under a close-lid or in an oven (see another paper of ours in Journal of Composite Material). In a LRI process performed in an oven, the resin temperature varies between 99°C-103°C. In this case, we have a quasi-isothermal test condition.
2. The thermo-chemical model is being developed (references 1 and 2) for upcoming curing and cooling stages modelling.
3. Using our current filling model but we can model properly the resin mass, fibre volume fraction and preform thickness. Consequently, we obtained good agreements in these parameters, but filling time (see tables 3 and 4). On another hand, we do have pointed out the importance of the temperature during filling stage (see table 6).

Now, the current paper shows the analyses of experimental approach and numerical simulation, and the general comparison of major parameters of the LRI process. Corresponding research work should be continued. The development of thermal model will be one of our important perspectives.

In figure 10, TC3-TC6 show the temperature of the preform, but not the one of the resin. Finally, a reference to another paper of ours connected with LRI test in an oven and some comments have been added in the section 4.2.1 to clarify.

4. The discussion on the linearity of viscosity and filling time (section 3.3.2) could be removed. As the isothermal filling simulation was used, we can predict this linearity easily in the numerical simulation as well as in the analytical solution.

Author's answer: The section 3.3.2 has been removed (Table 1 and Figure 5 have been removed relatively).

5. One of the important features in composite manufacturing processes employing flow channel (or high permeability layer) is the flow lead-lag effect. The flow in the high permeability layer leads the flow in the preform and the transverse flow through this lead-lag zone makes a significant contribution to the preform impregnation. Hence, through-thickness permeability is a key parameter to process modeling. In this work, the authors presented only the in-plane permeability (Eq. 8) in the numerical simulation. Then, the transverse permeability was estimated in the section 6.3. What value of the transverse permeability was used in the simulation results presented before the section 6.3? In Fig. 14, the curve for the permeability measured experimentally was provided. How was it obtained? By an independent measurement? Or by inverse identification? Please specify it.

Author's answer: Permeabilities are first modelled using a Carman-Kozeny's approach (Equation 6), corresponding to $5.10^{-14}$ $m^2$ in every direction. Then Equation 8 is used, it represents an analytical expression of the saturated transverse permeability measured by R. Nunez on UD fabric G1157. Influence of the permeability is addressed in section 6,3. Differences in using these measurements and Carman-Kozeny' approach are shown in Figure 13.

6. Darcy's law was introduced in Equation (3). If there is a movement of fiber (i.e. fabric deformation), fiber velocity should also be considered to compute Darcy's velocity (U_darcy-U_fiber, see the reference [20].). If the fiber velocity can be ignored, make some comment how this can be valid (e.g. dimensionless analysis).

Author's answer: It is one of the main strengths of our macroscopical approach [1, 2, 21] to accounted for resin flow in deformable preforms, since solid/porous mechanics is directly coupled with fluid mechanics. This can be represented using Updated Lagrangian scheme where fluid flows across the updated configuration, or using an ALE approach where convective (differential) velocity is accounted for in conservation equations.

7. The term "porosity" has been used as different meanings through the manuscript. In Page 5, "porosity" means 1-Vf." In page 17, "porosity" means "void type defects." Please rephrase "porosity" to be consistent.

Author's answer: It has been corrected, now the "porosity" means 1-Vf in the paper.

8. All the test conditions should be clearly described. For example, the authors presented in Tables 1-3, the thickness and the fiber volume fraction of the preform at the moment of "after infusion." In general, the thickness and the fiber volume fraction are not uniform through the part, just after "resin infusion." Were they obtained as the average values? Or, were they measured after the preform was relaxed after the fiber volume fraction became uniform during the post filling stage?

Author's answer: All the test conditions have been re-verified and described as clearly as possible. Tables 2 and 3 (in revised version) show the numerical simulation results. For the numerical simulation, the thickness of the preform stacking is uniform under our infusion assumption (the resin fills rapidly the draining fabric and then infuses the preform by and by through the thickness).

As pointed by reviewer, however, in the experimental study thickness is obviously not strictly uniform, as demonstrated in a paper of ours [25] to be published. Tables 3-6 (in revised version) have been modified to indicate the average (spatial) thickness.

In the section 4.1, the condition for preform compressibility test should be specified since the number of fabrics and the stacking sequence (i.e. nesting effect) may affect the test result.

Author's answer: To characterize the dry preform behaviour before resin infusion, an independent test of transverse compression with the same reference tissue UD used in the following LRI test, the number of fabrics and the stacking sequence have been added in the section 4.1.

9. English should be polished throughout the manuscript. I do not want to list all the writing issues, but just some representative examples.

- Awkward expressions to rephrase:

final piece (Page 1), industrial piece shapes (Page 1): piece -> part or product
thanks to (throughout the manuscript)
proposed to deal with the approach of injection processes (Page 3) -> proposed the approach to deal with…
strong experimental problems (Page 22) -> difficult

- Vague description:

as the preform and resin temperature (Page 1): preform? preform thickness? fiber volume fraction? Be specific!
permit to access (Page 2)
resin flow and cure are distinct (Page 2)
numerical coupling (Page 4)
in the porous (Page 5) -> in the porous medium

- Typos:

across the compressible performs (Page 2)
dry and wet performs (Page 6)
Dary's law (Page 9)
See Fig. 11 (Page 18) -> Fig. 8?

table 4 (Page 18) -> Table 4

- Plural and singular nouns:

the deformations of the preform (Page 3),
responses of the preform (Page 4),
isothermal conditions (Page 7),
the previous calculations (Page 10): the only one calculation was provided before this statement!

- Grammar issues or misuse:

a key parameter in (Page 6) -> a key parameter to
is composed into (Page 7) -> is decomposed into
some recent progress permit (Page 7) -> some recent progress permits
elements number (Page 9) -> number of elements
control the temperature change (Page 13) -> monitor or observe
is remains (Page 14)
enters into (Page 15), enters quickly into (Page 16)
much more time (Page 22) -> much longer time
model able to (Page 23) -> model is able to

Author's answer: The English problems mentioned here have been corrected in the paper and further improvements have been brought by an English specialist.
Authors thank the reviewer for taking time to help us in improving the quality.

## Reviewer: 2

Comments to the Author:

The authors did quite difficult experiments. But, it is not easy to find a new idea or experimental results. Most of experimental results can be deduced in an engineering sense by another researchers.

**P.14**
In LRI process, a pressure applied in resin is equal or lower than atmospheric pressure. Therefore, Cauchy stress (Fig.7) obtained in the experiment of force versus displacement through the thickness of the preform is quite high compared to normal resin pressure of LRI process.

Author's answer: Yes, in LRI process, the pressure applied in resin normally equates atmospheric pressure. Figure 7 presents the experimental compression curve in out-of plane direction for dry UD fabric G1157 upto a fiber volume fraction of 70%. Here, we want to show that the dry fabrics used in our resin infusion test have a strongly non-linear behaviour. However, the beginning of this compression curve (Cauchy stress < 1E+05 Pa) is useful in the numerical simulations to follow the finite deformations.

**P.16**
It should be described whether the viscosity of RTM6 was measured in this study or was given by the manufacturer.

Author's answer: The viscosity of RTM6 is not measured in this study but given by the manufacturer (HEXCEL). This explication has been added in the paper (in the section 4.2.1).

**P.20**
In table 5, fiber volume fraction after filling is the same 56% at the experiment and simulation. But mass of resin used during the infusion stage are different, respectively 350g and 375g. I think the used resin is dependent on final fiber volume fraction instead of initial fiber volume fraction. The author should explain the reason.

Author's answer: Reviewer 2 is perfectly right, a further precision has been added : 56% is a rounded number. More precise fiber volume fractions after filling are now given : 55.5% and 56.1% respectively for the experiment and the numerical simulation. These fiber volume fractions depend highly on the thickness of the preform.

Tables 3, 4, 5 and 6 (in revised version) have been modified accordingly.